%% file: main.tex
\documentclass[onecolumn]{aastex62}

\usepackage{comment}
\usepackage{hyperref}

\begin{document}

\author{Julia Falcone}
\affiliation{Department of Physics and Astronomy, Georgia State University, Atlanta, GA 30303, USA}
\title{A Simple Method of Producing Images of SDSS Spectra in a Free Spreadsheet Program}

\input{sec1_abstract.tex}

\input{sec2_intro.tex}

\input{sec3_methods.tex}

\input{sec4_troubleshoot.tex}

\input{sec5_ack.tex}

\input{sec6_references.tex}

\nocite{*}
\bibliographystyle{aasjournal} 

\end{document}

%% file: sec1_abstract.tex
\section{Abstract}
Using Google Sheets, I develop a method to easily reproduce thousands of images of SDSS spectra so that they may be studied in only a fraction of the time it would otherwise take. This method may be helpful in projects requiring large samples of SDSS objects with spectra, and is described in a step-by-step manner so that it is accessible to everyone. 

%% file: sec2_intro.tex
\section{Motivation}

As of the fourth generation of the Sloan Digital Sky Survey (SDSS-IV), Data Release 15 (DR15) \citep[]{blanton, aguado}, over 2 million galaxies have been observed with single-fiber spectra which visualize flux levels over a wide range of wavelengths\footnote{\href{sdss.org/dr15/scope}{sdss.org/dr15/scope}}. These spectra are invaluable in projects involving the observation of properties like broad-line emission or characteristics of specific galaxy morphologies. However, when projects require one to look through thousands of spectra to obtain a sufficient sample size, the process of clicking thousands of different links to arrive at the images of desired spectra can be extremely time-consuming. There have certainly been attempts to automate morphological classifications  \citep[e.g.][]{sreejith, dediego}, but others have shown that human observation of spectra may yield more accurate results, especially when classifications rely on noticing subtle spectral characteristics that are difficult for automated systems to spot \citep[]{greene}. In this research note, I detail a method to easily and quickly reproduce images of SDSS spectra in Google Sheets\footnote{\href{google.com/sheets/about}{google.com/sheets/about}}, which allows for thousands of spectra to be observed by the human eye in only a fraction of the time it would normally take, by centralizing all of the necessary data in one place instead of over a multitude of webpages. This method was developed when my research group was tasked with looking through more than 15,000 spectra to identify E+A galaxies \citep[Liu et al. 2020, in preparation;][]{dresslergunn} in numerous large fields. This technique reduced the time to complete the assignment by at least a factor of 10.

%% file: sec3_methods.tex
\section{Methodology}

I have written a Github post\footnote{\href{juliafalcone.github.io/googlesheets-supplement}{juliafalcone.github.io/googlesheets-supplement}} explaining the steps in this section steps in greater detail for anyone who might find it useful. As mentioned, the application I find best suited for this method is Google Sheets, which allows for the extraction of images from websites with the correct use of its library of functions. The only necessary information for each galaxy is its spectral ID (hereafter called the SpecID), RA, and declination, all of which can be obtained when sending a data query through the SDSS SkyServer Search Form\footnote{\href{skyserver.sdss.org/dr16/en/tools/search/form/searchform.aspx}{skyserver.sdss.org/dr16/en/tools/search/form/searchform.aspx}} and checking the applicable boxes. The RA and dec will only be necessary in the event of troubleshooting (Section \ref{sec4}). 

Use Figure \ref{fig:graphs} as a reference where the first, second, and third columns contain each galaxy's RA, dec, and SpecID respectively. Then, if the fourth column is to display an image of the galaxy's spectrum, that column's cell for a given row should contain the following text:
\vspace{2mm}

\texttt{=IMAGE(CONCATENATE("http://skyserver.sdss.org/dr15/en/get/SpecById.ashx?id=",[SpecID]),1)} 
\vspace{2mm}

in which \texttt{[SpecID]} should point to the cell containing that galaxy's SpecID. This command uses \texttt{CONCATENATE} to combine the SpecID with an otherwise static URL to produce a link of the spectrum's image. The \texttt{IMAGE} function grabs the image of the spectrum from that URL and inserts it into the cell as shown in Figure 1. The second parameter in the \texttt{IMAGE} function (in this example, \texttt{1}) relates to the different options in displaying the image which is detailed in the function's description\footnote{\href{support.google.com/docs/answer/3093333}{support.google.com/docs/answer/3093333}}. 

This produces the image of the spectrum in the desired cell, and can then be applied to all other rows.

%% file: sec4_troubleshoot.tex
\section{Troubleshooting} \label{sec4}

Since this method makes use of the functions available in Google Sheets, it is not applicable to other similar applications such as Excel. Even in Google Sheets, this method produces null results approximately 15 percent of the time, as the SpecID does not always correspond with the galaxy with which it is listed. In such cases, the correct SpecID can be obtained using one of the following solutions:
\vspace{5mm}

    \textbf{Problem}: After inputting the command above, I receive an error message.
    
    \textbf{Solution}: First, make sure that the \texttt{[SpecID]} text is deleted and replaced with the correct cell containing the SpecID. Additionally, depending on how the command is copied and pasted, there may be inconsistencies in how the quotation marks are represented. Make sure that all quotation marks are of the Unicode number U+022(\texttt{"}) and not the left and right double quotation marks U+201C/U+201D (“ "). 
  \vspace{5mm}  
    
    \textbf{Problem}: After inputting the command above, the cell reads “No image exists for this region."
    
    \textbf{Solution}: When this happens, it means the SpecID which was received from the SkyServer query is incorrect. To remedy this problem, one can either obtain it manually by searching through SDSS using the RA and dec, or one can copy the following command into the cell containing that galaxy's SpecID:
    \vspace{2mm}
  
    \texttt{=MIDB(INDEX(IMPORTHTML(CONCATENATE("http://skyserver.sdss.org/dr15/en/tools/explore/Summary}
    
    \texttt{.aspx?ra=",[ra],"\&dec=",[dec]), "table",28),1),14,19)}
    \vspace{2mm} 
    
    Here, \texttt{[ra]} and \texttt{[dec]} are the cells containing the galaxy's RA and dec, respectively. This command makes use of several other functions, which have the following purposes: \texttt{CONCATENATE} once again creates a URL from inserted values of the galaxy's RA and dec; \texttt{IMPORTHTML} imports the data from this URL into the sheet in the form of a table (which I have specified with the parameter \texttt{"table"}). The SpecID is typically on row 28 of this table, so I choose to isolate that row; \texttt{INDEX} further helps to isolate the SpecID; and lastly \texttt{MIDB} is used to pinpoint the exact location of the desired characters. The SpecID is the 14th character in the row and runs 19 digits long, hence the final two numbers. Although this command stretches over two lines in this note, it should read as one continuous line in the program, so it will be necessary to delete any line breaks once the command is pasted into the program.

\vspace{5mm}

    \textbf{Problem}: I typed in the command from the solution above to obtain the SpecID, but the cell for the spectrum is now completely blank.
    
    \textbf{Solution}: Change  \texttt{28} to  \texttt{29} in the command above, so it reads as such:
  \vspace{2mm}
  
    \texttt{=MIDB(INDEX(IMPORTHTML(CONCATENATE("http://skyserver.sdss.org/dr15/en/tools/explore/Summary}
    
    \texttt{.aspx?ra=",[ra],"\&dec=",[dec]), "table",29),1),14,19)}
     \vspace{2mm}
 
    The problem is caused by flags on some galaxies warning of possibly unreliable photometry. The inclusion of this text causes the placement of all other measurements below it to shift by a line.

%% file: sec5_ack.tex
\section{Acknowledgements}

I thank Charles Liu for his valuable discussions and input. This work was supported by the Alfred P. Sloan Foundation via the SDSS-IV Faculty and Student Team (FaST) initiative, ARC Agreement SSP483, and by NSF grant AST-1460860 to the CUNY College of Staten Island.

%% file: sec6_references.tex

\bibliographystyle{plain}
\bibliography{bibbo.bib} 

\begin{figure}    
\includegraphics[width=\textwidth]{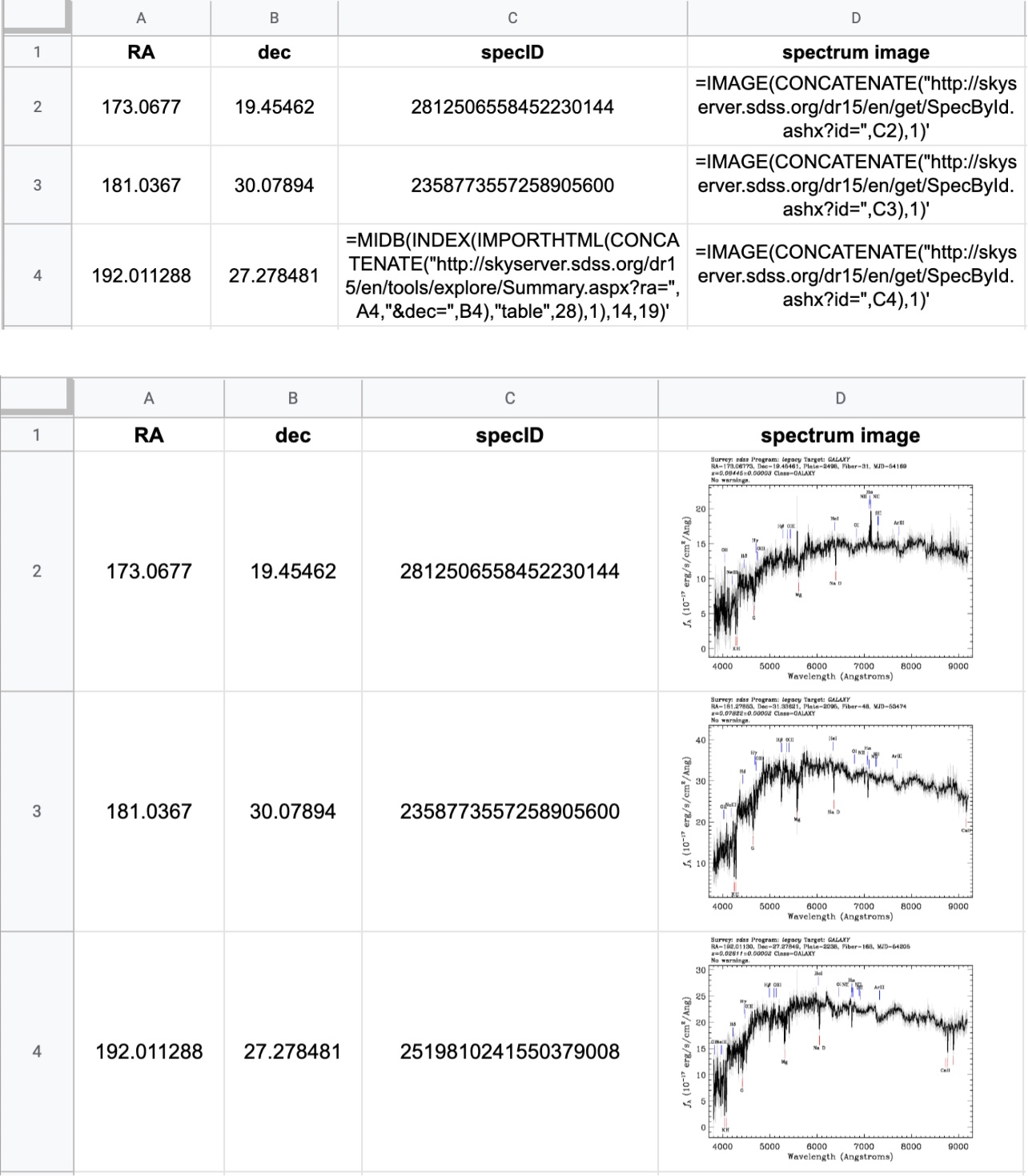}
\caption{Above: An example table on Google Sheets containing the commands necessary to produce the desired images. Whereas the first two rows have the correct SpecID, the third row does not and commands are entered into the cell as described in Section 3. Bottom: The resulting table after the commands are entered, showing images of spectra inside the cells corresponding to each SpecID.}
\label{fig:graphs}
\end{figure}